\documentclass[PRL,aps,twocolumn,showpacs,preprintnumbers,amsmath,amssymb,superscriptaddress,nofootinbib]{revtex4-1}
\usepackage[dvips]{graphicx}
\usepackage{epsf}
\usepackage{amsmath}
\usepackage{amssymb}
\usepackage{amsfonts}
\usepackage{times}
\usepackage[usenames]{color}
\usepackage[normalem]{ulem}
\usepackage[T1]{fontenc}
\usepackage[dvipsnames]{xcolor}
\usepackage{floatrow}

\usepackage{graphicx}
\usepackage{dcolumn}
\usepackage{bm}
\usepackage{color}

\begin{document}

\title{An accurate method to determine the systematics due to the peculiar velocities of galaxies in measuring the Hubble constant from gravitational wave standard sirens}

\author{Jian-hua He}
\email[Email address: ]{hejianhua@nju.edu.cn}
\affiliation{School of Astronomy and Space Science, Nanjing University, Nanjing 210093, P. R. China}
\affiliation{Institute for Computational Cosmology, Department of Physics, Durham University, South Road, Durham DH1 3LE, UK}
\begin{abstract}
We propose a novel approach to accurately pin down the systematics due to the peculiar velocities of galaxies in measuring the Hubble constant from nearby galaxies in current and future gravitational-wave (GW) standard-siren experiments. Given the precision that future GW standard-siren experiments aim to achieve, the peculiar velocities of nearby galaxies will be a major source of uncertainty. Unlike the conventional backward reconstruction that requires additional redshift-independent distance indicators to recover the peculiar velocity field, we forwardly model the peculiar velocity field by using a high-fidelity mock galaxy catalog built from high-resolution dark matter only (DMO) N-body simulations with a physically motivated subhalo abundance matching (SHAM) technique without introducing any free parameters. Our mock galaxy catalog can impressively well reproduce the observed spectroscopic redshift space distortions (RSDs) in highly non-linear regimes down to very small scales, which is a robust test of the velocity field of our mock galaxy catalog. Based on this mock galaxy catalog, we accurately, for the first time, derive the peculiar velocity probability distributions for the SDSS main galaxy samples. We find that the systematics induced by the peculiar velocities of SDSS like galaxies on the measured Hubble constant can be reduced to below $1\%$($1\sigma$) for GW host galaxies with a Hubble flow redshift just above $0.13$, a distance that can be well probed by future GW experiments and galaxy surveys. 
\end{abstract}

\maketitle
\section{Introduction}
The coeval detection of gravitational waves (GW170817)~\cite{TheLIGOScientific:2017qsa} and a $\gamma$-ray burst (GRB 170817A)~\cite{Goldstein:2017mmi} from the merger of two neutron stars ushered us in a new era of astronomy. This is the first time that we are able to study the Universe with both vision and hearing. This multi-messenger observations of both gravitational wave (GW) and their electromagnetic (EM) counterparts allow us to use gravitational wave (GW) as standard sirens~\cite{Nature1986}: the intrinsic total gravitational luminosity can be derived to unprecedented precision from the precise way in which GW evolves, given the much higher sensitivity of the future ground- and space-based GW experiments such as the Einstein Telescope (ET)~\cite{Einstein_tele}, 40-km LIGO~\cite{LIGO40}, eLISA~\cite{eLISA}, and DECIGO~\cite{Sato_2009}; then coupled to the measured absolute strain amplitude of GW, the luminosity distance $D_{cos}$ to the source can be accurately determined as well; finally, from the EM counterpart, a unique host galaxy can be identified, which makes it possible to obtain a spectroscopic redshift follow-up $z_{obs}$. As such, GW provides a new technique to measure the Hubble constant from nearby galaxies (redshift $z\ll1$)
\begin{equation}
H_0^{obs}=\frac{cz_{obs}}{D_{cos}}\,,\label{Hobs}
\end{equation}
where $c$ is the speed of light in vacuum. An advantage of this technique is that it is based on the first principles of general relativity rather than the empirical scaling relations that are widely used in the conventional astronomical distance indicators.

This independent way of getting cosmic distances is of particular importance, given the current tension between the value of the Hubble constant determined by Type Ia supernovae via the local distance ladder $73.21\pm0.74$~\cite{Riess:2016jrr} and that from Cosmic Microwave Background observations ($66.93\pm0.62$)~\cite{Aghanim:2018eyx}. Although significant effort has been made to investigate this tension, there is no obvious systematic origin reported~\cite{Riess:2016jrr,Aghanim:2018eyx}. The true reason behind this tension is still elusive. Therefore, a more effective way of measuring the Hubble constant is urgently called for. 

\begin{figure}
\includegraphics[width=\linewidth]{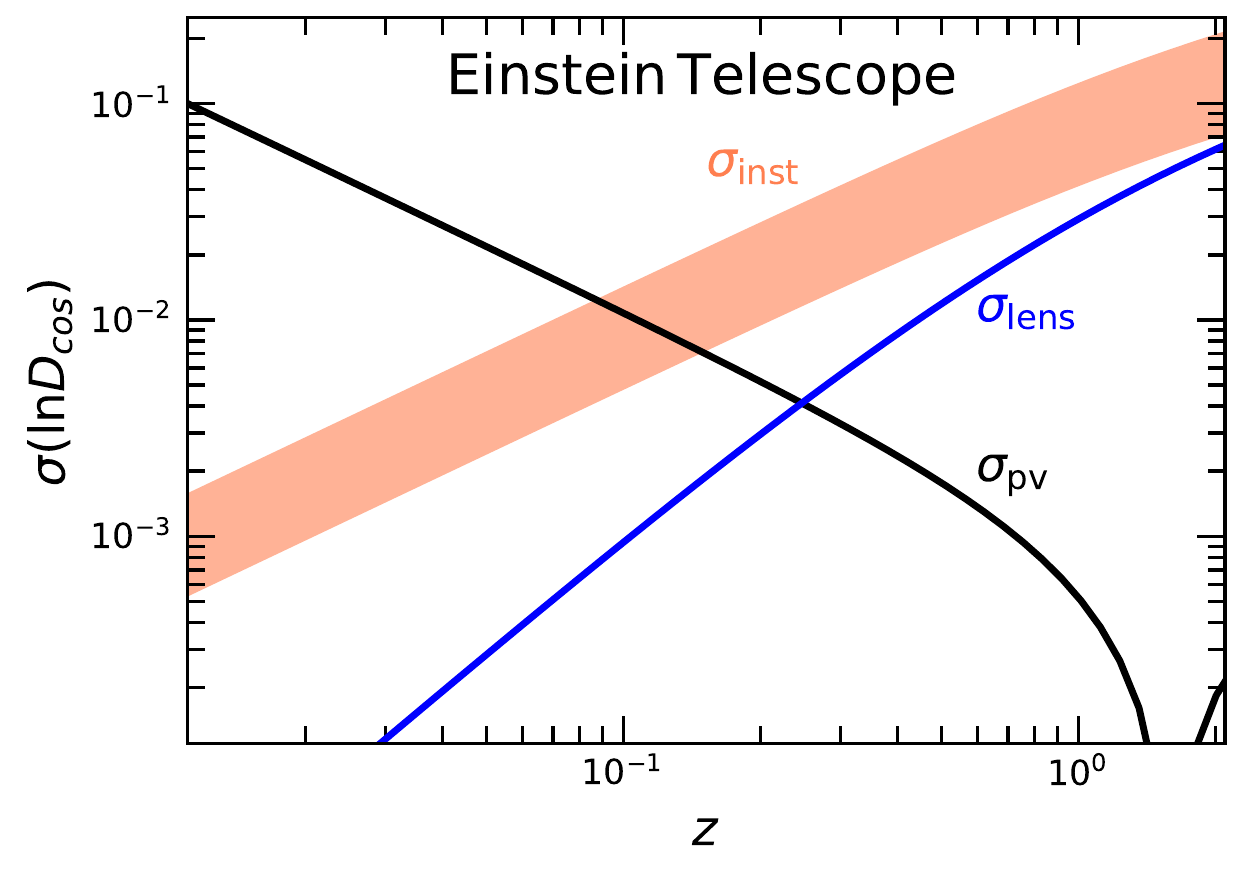}
\caption{The predicted errors on the luminosity distance as a function of redshift for the Einstein Telescope, a third-generation ground-based observatory. Here, we assume a single GW source that is composed of two neutron stars with $1.44$ solar mass each. The shaded region represents the estimated range of errors due to the instrumental noise and the geometry of the source relative to the telescope, such as the location of the source on the sky ($\theta\,,\phi$), the polarization angle $\psi$, and the inclination of the binary's angular momentum relative to the line-of-sight direction $\iota$. We assume $\iota<20^{\circ}$ and treat $\theta\,,\phi\,,\psi$ as free parameters. The shaded region is estimated using random sampling with a uniform distribution on the sky and the polarization angle. The solid black line is the error due to the galaxy peculiar velocities and the blue line is the error due to the lensing effect. Note that $\sigma_{\rm inst}$ scales with the number of GW events as $1/\sqrt{N}$. With $1000$ events, $\sigma_{\rm inst}$ can be reduced by a factor of $31.6$. Therefore, at low redshift, galaxy peculiar velocity is the dominant source of uncertainty.  \label{error}}
\end{figure}

\section{Systematics of GW standard sirens}At high redshift, the systematics of GW standard sirens are dominated by the instrumental uncertainty of GW telescopes as well as the effect of lensing~\cite{PhysRevD.83.084045}. The instrumental error $\sigma_{\rm inst}$ is sensitive to the telescope noise power spectral density, which determines the detectability of the telescope. Due to the pattern of the antenna, the sensitivity of the telescope also depends on the sky position ($\theta\,,\phi$). On the source side, the inclination of the binary's angular momentum relative to the line-of-sight direction $\iota$ also affects the observed GW signal. Following Ref.~\cite{Zhao:2010sz}, we assume that the source of GW comes from two neutron stars with $1.44$ solar mass each, and the EM counterpart is a short gamma-ray burst, a highly beamed phenomenon, that can put constraints on the inclination angle $\iota<20^{\circ}$. We use the Fisher information matrix to estimate the error on the luminosity distance, following the framework presented in Ref.~\cite{Finn:1992wt}. In our analyses, we treat the sky position ($\theta\,,\phi$), and the polarization angle $\psi$ as free parameters and use random sampling with a uniform distribution on the sky and the polarization angle $\psi$ to estimate the possible range of errors due to these free parameters. In addition, we estimate the error due to the lensing effect by~\cite{GWlensing}
\begin{equation}
\sigma_{\rm lens}(z)=0.066\left[\frac{1-(1+z)^{-0.25}}{0.25}\right]^{1.8}\quad.
\end{equation}
The shaded region in Fig.~\ref{error} shows the predicted range of the instrumental error on the luminosity distance for a single GW event. Here, we focus on the Einstein Telescope, a third-generation ground-based observatory, for illustrative purposes. The solid blue line shows the error due to the lensing effect.

Aside from the instrumental noise of the GW telescopes, at low redshift another major source of uncertainty comes from the peculiar motions of nearby galaxies. In addition to the Hubble flow, a galaxy can acquire an additional redshift $z_{pec}$ due to its peculiar motion. The observed redshift $z_{obs}$ is indeed related to the Hubble flow redshift $z_{cos}$ by~\cite{Harrison,Davis:2014jwa}
\begin{equation}
1+z_{obs}=(1+z_{pec})(1+z_{cos})\,.\label{redshiftrelation}
\end{equation}
If the velocity of a galaxy is non-relativistic, the peculiar redshift $z_{pec}$ is related to the peculiar velocity $v_{pec}$ by $z_{pec}=v_{pec}/c$. Equation~(\ref{redshiftrelation}) then gives
\begin{equation}
z_{obs}=z_{cos}+\frac{v_{pec}}{c}(1+z_{cos})\,.\label{zobs}
\end{equation}
Inserting Eq.~(\ref{zobs}) into Eq.~(\ref{Hobs}), we obtain
\begin{equation}
\frac{H_0^{obs}-H_0^{cos}}{H_0^{cos}}=\frac{v_{pec}}{c}\frac{(1+z_{cos})}{z_{cos}}\,.\label{H0measure}
\end{equation}
where we have used $H_0^{cos}=cz_{cos}/D_{cos}$.
From Eq.~(\ref{H0measure}), even without the instrumental error $\sigma_{\rm inst}$, the accuracy of measuring the Hubble constant is still limited by the peculiar velocity field $v_{pec}$. In a virilized galaxy cluster, $v_{pec}$ can be as large as $\sim500{\rm km/s}$. The uncertainty induced by such high-speed motion is substantial. In Fig.~\ref{error}, we estimate such error on the luminosity distance by~\cite{Gordon:2007zw}
\begin{equation}
\sigma_{\rm pv}(z)= \mid1 -\frac{(1+z)^2}{H(z)D_{cos}(z)}\mid\sigma_{v}\quad,
\end{equation}
where we set $\sigma_{v}=331 {\rm km/s}$. The choice of this value will be discussed in detail later. From Fig.~\ref{error}, clearly, at low redshift, galaxy peculiar velocity is the dominant source of uncertainty. Further note that, the instrumental error $\sigma_{\rm inst}$ scales with the number of GW events as $1/\sqrt{N}$, which can be significantly reduced in the future, given the number of GW events future GW experiments aim to detect. Moreover, combining different detectors, future advanced GW detector network can provide better measurements of the source inclination and polarization, which can break the degeneracy between distance and inclination, and, therefore, further improve the accuracy of the distance measurement. However, the error due to the galaxy peculiar velocity $\sigma_{\rm pv}(z)$ , on the other hand, is intrinsic, which can not be easily mitigated. Therefore, to pin down the velocity field $v_{pec}$ is crucial to accurately determine the Hubble constant in the future GW experiments from nearby galaxies~\cite{Wang:2017jum}.

\section{SHAM mock catalog} However, directly determining $v_{pec}$ is indeed non-trivial. Usually, a secondary redshift-independent distance indicator has to be used, such as the Tully-Fisher (TF) relation~\cite{TFrelation} or the Fundamental Plane (FP) relation~\cite{FP1,FP2}. The former is a scaling relation for late-type galaxies that express the luminosity as a power law function of rotation velocity and the latter is a scaling relation for elliptical galaxy spheroids-bulk motions. These two relations can provide the largest number of distance measurements for galaxies, from which it is possible to reconstruct the bulk motion of galaxies at large scales. However, this method has several limitations: the distance indicators only work for particular types of galaxies. For some types of galaxies, their distances cannot be obtained in this way. The reconstructed peculiar velocity fields may also suffer the selection effect, namely, the reconstructed values depend heavily on the samples used. This is due to the fact that different types of galaxies intrinsically move differently. Moreover, in fact, in addition to the coherent bulk motion at large scales, galaxies also have local random motions. For galaxies in a virilized galaxy cluster, random motions are the dominated factor to the peculiar velocity rather than the bulk motion at large scales. The highly non-linear nature of such random motions poses a challenge to determine the peculiar velocity fields accurately.

In order to effectively overcome these difficulties, here we propose a forward-modeling method. Rather than backward reconstructing the peculiar velocity field, we forwardly model the peculiar velocity field using a high fidelity mock galaxy catalog that is built from high-resolution N-body simulations, from which the peculiar velocity field can be explored to very non-linear regimes.  

\begin{figure*}
\includegraphics[width=\linewidth]{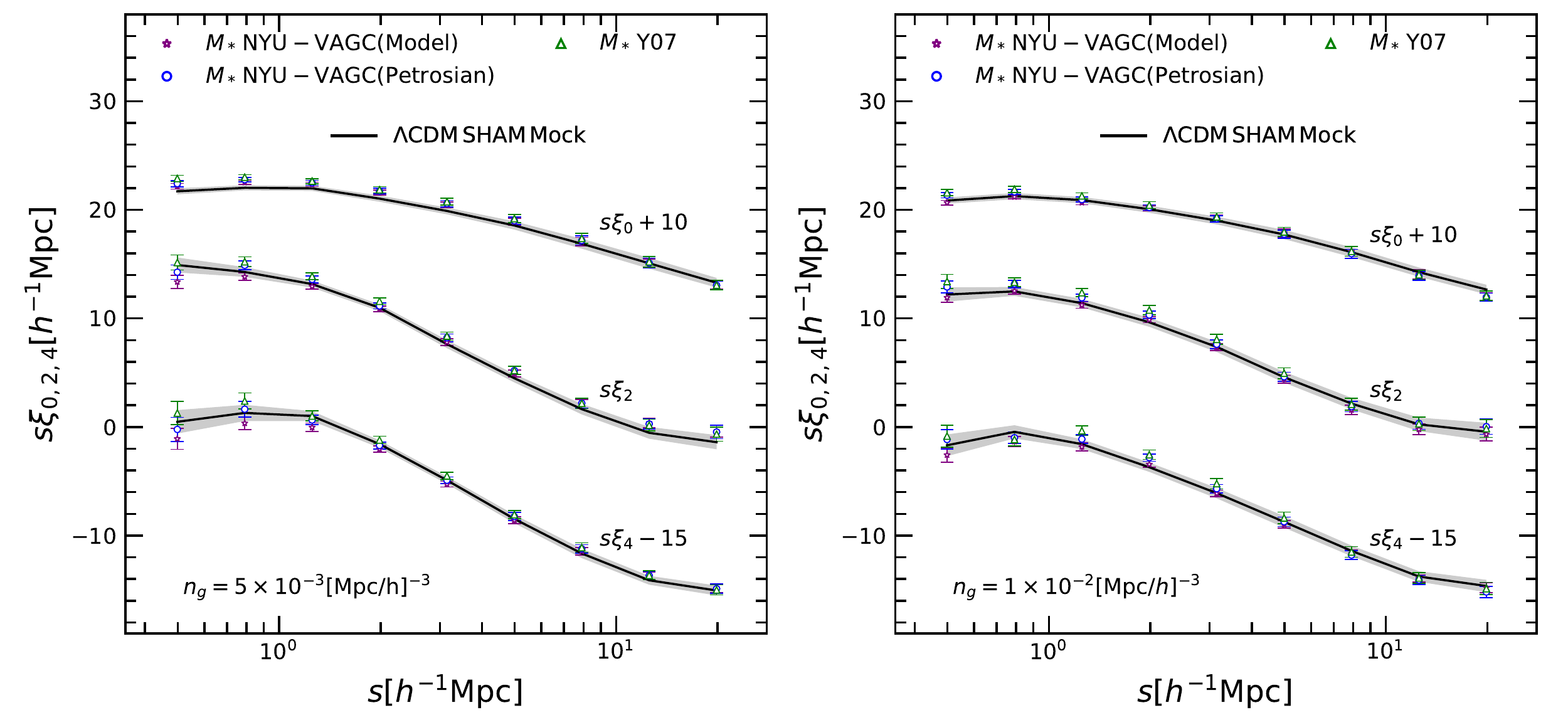}
\caption{Redshift-space multipoles (monopole $\xi_0$, quadrupole $\xi_2$, and hexadecapole $\xi_4$) for our SHAM mock galaxy catalog (solid black lines) and the measurements from the SDSS main galaxy samples (symbols with error bars). Overall, the predictions of our SHAM mock can match the observations impressively well down to very small scales. Left: galaxy samples with a low number density $n=0.005[{\rm Mpc}/h]^{-3}$. Right: the same but for galaxy samples with a high number density $n=0.01[{\rm Mpc}/h]^{-3}$. The low-density sample $n=0.005[{\rm Mpc}/h]^{-3}$ has more total number of objects due to its relatively large volume and, therefore, is more robust in the two-point statistics, while the high-density sample $n=0.01[{\rm Mpc}/h]^{-3}$ is more robust against various systematics. In order to demonstrate the robustness of our RSD measurements, we show three different estimators of stellar masses: a template-fit method as adopted in the NYU catalog with the SDSS model magnitudes (stars), the same template-fit method but with SDSS Petrosian magnitudes (circles), and a single-color method (triangles). The error bars are derived from 133 realizations using the jack-knife re-sampling technique. The black shaded regions represent the 1$\sigma$ uncertainty in the theoretical predictions.\label{RSD}}
\end{figure*}

Usually, mock galaxy catalogs are constructed using phenomenological frameworks, such as the halo occupation distribution (HOD)~\cite{Jing:1997nb,Peacock:2000qk,Berlind:2001xk,Zheng:2004id} or the conditional luminosity function (CLF)~\cite{Yang:2002ww,Yang:2008eg}. The basic idea of this approach is that galaxies reside in dark matter halos. And the probability of the distribution of galaxies is only dependent on the masses of dark matter halos. However, these approaches neglect some important effects such as the assembly bias~\cite{Gao:2005ca,Gao:2006qz}. And they are also shown to be difficult to reproduce the observed small scale RSDs unless additional velocity bias is added~\cite{Guo:2015epa,Guo:2015dda}. Therefore, the velocity field derived from these mock catalogs cannot be directly used to pin down the systematics in measuring the Hubble constant.

Rather than using the frameworks of HOD, we instead use the subhalo abundance matching technique~\cite{Kravtsov:2003sg,Vale:2004yt,Conroy:2005aq,moster,qiguo,Reddick_2014}. This approach assumes that there is a one to one correspondence between a property of galaxy and that of a dark matter halo. The property of the galaxy is identified as its stellar mass (mass in stars), while the property of the dark matter halo is identified as the peak value of its maximum circular velocity during its merger history $v_{\rm peak}$. These choices are strongly motivated by state-of-the-art hydro-dynamical simulations of galaxy formation~\cite{Chaves-Montero:2015iga}, from which these two properties are shown to be the most tightly correlated with each other. Moreover, in our implementation, we do not add any scatter between $v_{\rm peak}$ and galaxy stellar mass. So there is no free parameter in our approach. The advantage of our approach is that it is completely determined by DMO simulations which are further based on the first principle of gravity. So assuming GR-$\Lambda$CDM, our approach provides a method that relies only on first principle calculations to recover the underlining velocity fields. 

Our SHAM mocks can be directly compared to observations. On the observation side, we use the Sloan Digital Sky Survey (SDSS) main galaxy sample. Specifically, we adopt the New York University Value-Added Galaxy Catalog (NYU-VAGC) [26], which is an enhanced version of 7th Data Release of the SDSS main galaxy sample. The catalog covers an area of $7732\,{\rm deg}^2$, which are mainly located in a contiguous region in the north Galactic cap. We also include the three strips in the south Galactic cap. Based on this catalog, we construct volume-limited samples that are complete in galaxy stellar mass (see Ref.~\cite{He:2018oai} for details). On the theory side, our SHAM mock catalog is based on the Small MultiDark Planck simulation (SMDPL)~\cite{Klypin:2014kpa}. We build a full-sky mock by collating 8 replicas of the box and place the observer at the center. Redshift distortion effects are obtained for each galaxy by projecting its velocity along the line-of-sight to the observer. We also add exactly the same survey mask as the real data in our mock galaxy catalog. 

In order to mitigate the uncertainty in the estimate of galaxy stellar mass, we use galaxy number densities instead of applying a stellar mass threshold to select our galaxy samples. The idea is to keep the rank-order of galaxies stable. As shown in Ref.~\cite{He:2018oai}, a higher number density can also help to further mitigate various systematics. Therefore, we impose a minimum number density on our galaxy samples $n\ge0.005[{\rm Mpc}/h]^{-3}$. In addition, we only consider the volume-limited samples that have a reasonably large volume in order to mitigate the impact of the cosmic variance and the impact of missing long-wave modes on the two-point statistics in galaxy clustering analysis. For this, we only use volume-limited samples that have the largest radial distance of the volume along the light-of-sight direction satisfying $r_{\rm max}\ge 200{\rm Mpc}/h$. Based upon the above considerations, our final choices are $n=0.005[{\rm Mpc}/h]^{-3}$ and $n=0.01[{\rm Mpc}/h]^{-3}$. The low-density sample $n=0.005[{\rm Mpc}/h]^{-3}$ has more total number of objects due to its relatively large volume and, thereby, is more robust in the two-point statistics, while the high-density sample $n=0.01[{\rm Mpc}/h]^{-3}$ is more robust against the impact of various systematics. 

We expand the redshift space two-point correlation function $\xi(r_{\sigma},r_{\pi})$ in terms of Legendre polynomials
\begin{equation}
\xi_l(s)=\frac{2l+1}{2}\int_{-1}^{1}d\mu \xi(s,\mu)P_l(\mu)\, ,\label{Intxi}
\end{equation}
where $P_l(\mu)$ is the Legendre polynomial of order $l$, $s=\sqrt{r_{\sigma}^2+r_{\pi}^2}$ and $\mu=r_{\pi}/s$. $r_{\sigma}$ and $r_{\pi}$ are the separations of galaxy pairs perpendicular and parallel to the line-of-sight direction, respectively. Figure~\ref{RSD} shows the predicted multipoles (monopole $\xi_0$, quadrupole $\xi_2$, and hexadecapole $\xi_4$) of RSDs (black solid lines) compared to the measurements from the SDSS data (symbols with error bars). The left panel is for the low-density sample $n=0.005[{\rm Mpc}/h]^{-3}$ and the right panel is for the high-density one $n=0.01[{\rm Mpc}/h]^{-3}$. The predicted RSDs from our mock galaxy catalog agree impressively well with the observations down to very small scales $r_{s}\sim 0.5{\rm Mpc}/h$, which is true for both the number densities considered. In addition, we also demonstrate the robustness of our RSD measurements, using three different estimators of stellar masses: a template-fit method originally adopted in the NYU catalog with the SDSS model magnitudes (stars)~\cite{Blanton:2004aa}, the same template-fit method but using SDSS Petrosian magnitudes (circles), and a single-color method (triangles)~\cite{Yang:2007yr}. From Fig.~\ref{RSD}, different stellar mass estimators give very similar results. Since RSDs, especially the small scale RSDs, are very sensitive to the motion of galaxies, they provide a robust test of the galaxy velocity field of our mock catalog. 

\section{Forecast of the accuracy on the Hubble constant} Based on our high fidelity mock galaxy catalog, we are able to accurately measure the velocity field of galaxies. Note that in our mock catalog, the velocity field of galaxies is directly taken from the velocities of dark matter subhalos, which are, in turn, based on the first principle calculations in the DMO simulations. Our results, therefore, are only dependent on the gravity model assumed.
 
\begin{figure}
{\includegraphics[width=\linewidth]{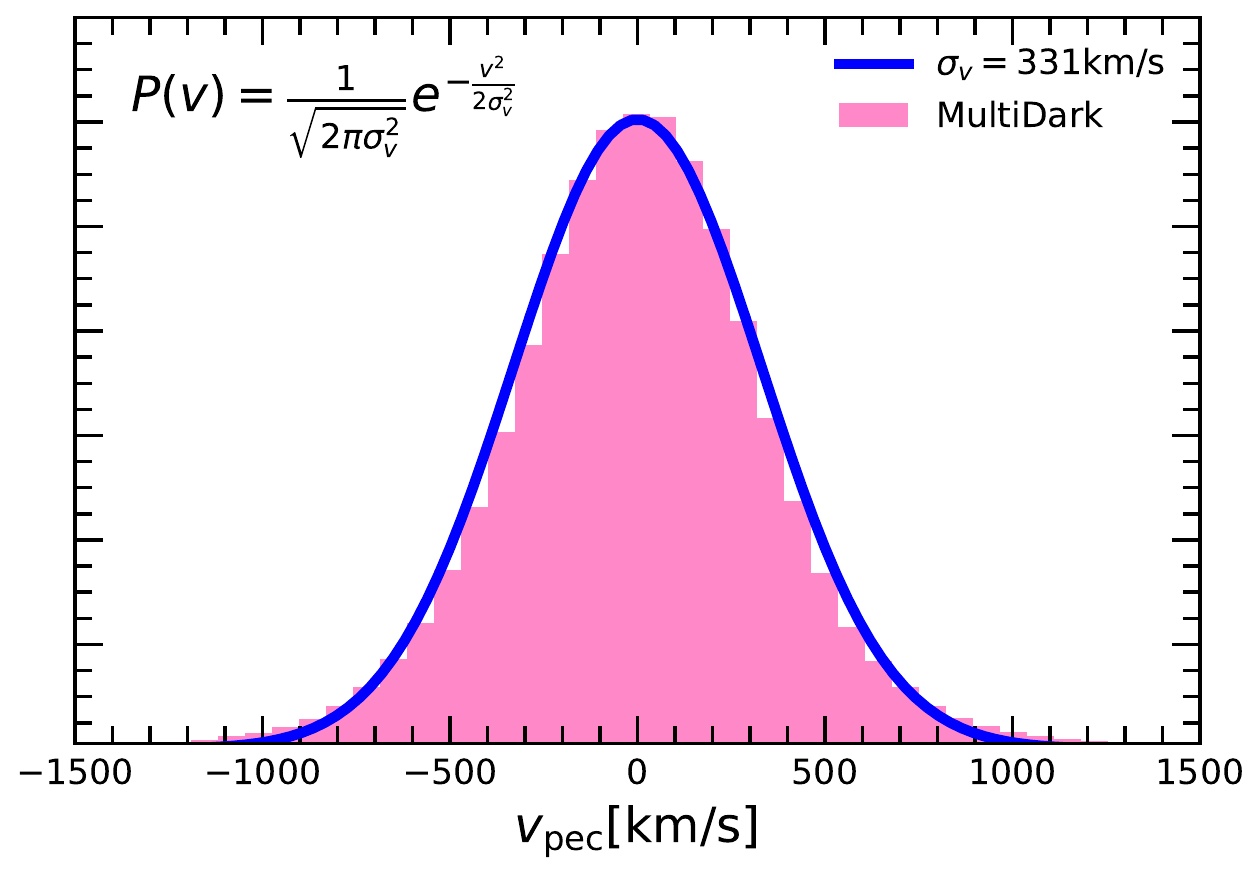}}
\caption{The peculiar velocity probability distributions along the light-of-sight for our mock galaxy catalog. The sample used here has a number density of $n=0.01[{\rm Mpc}/h]^{-3}$. The solid blue line represents a Gaussian fit. The velocity field can be reasonably well described by the Gaussian distribution except the extreme high speed wings. From this Gaussian fit, we derive the velocity dispersion as $\sigma_v=331{\rm km/s}$. \label{Vdis}}
\end{figure}

\begin{figure}
{\includegraphics[width=\linewidth]{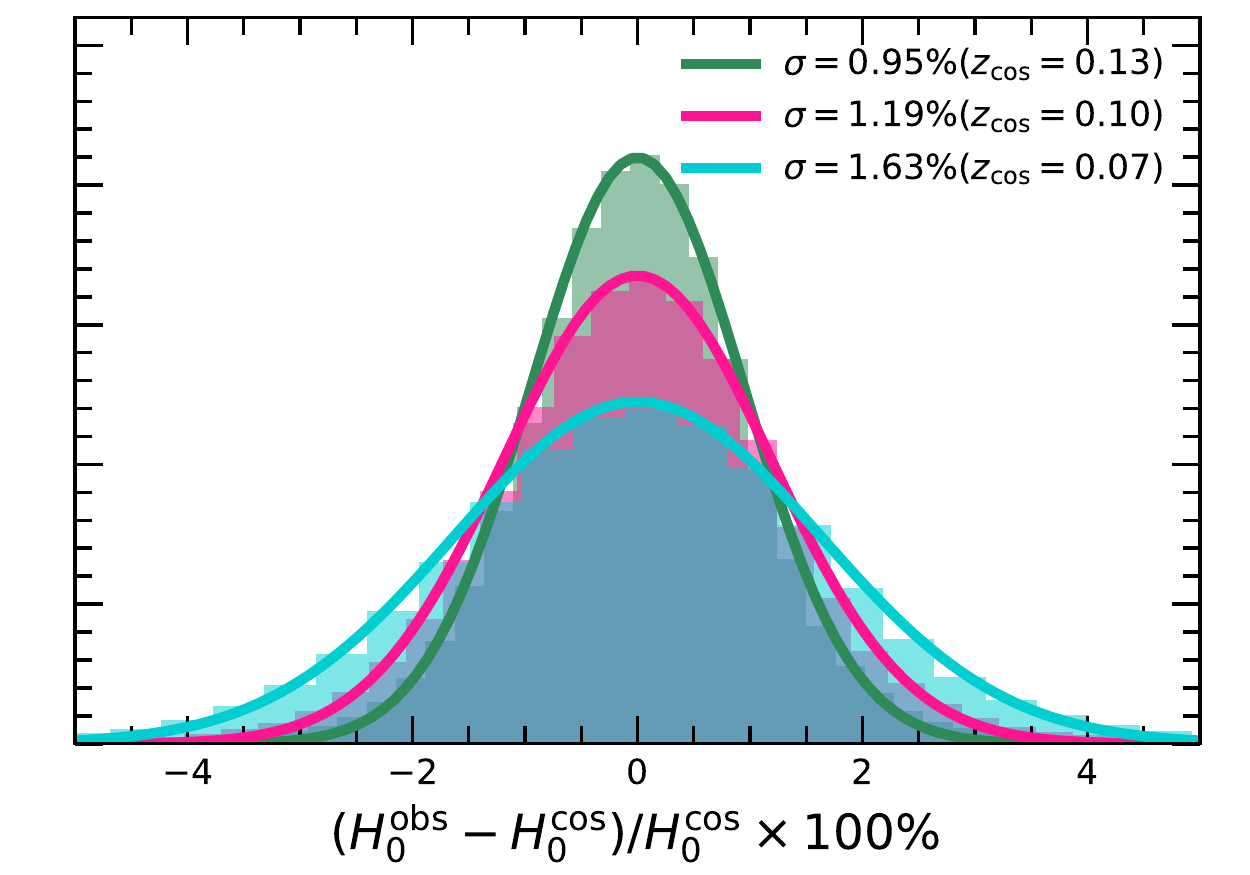}}
\caption{The probability distribution of the systematic errors on the Hubble constant due to the peculiar velocities of galaxies. Different colors represent the results for GW host galaxies with different Hubble flow redshifts $z_{cos}$. The solid lines give Gaussian fits to the errors. For GW host galaxies with  Hubble flow redshifts just above $z_{cos}=0.13$, the systematic errors can be reduced to below $1\%$. \label{sys}}
\end{figure} 
 
Figure~\ref{Vdis} shows the peculiar velocity probability distributions along the light-of-sight for our mock galaxy catalog for a galaxy sample with $n=0.01[{\rm Mpc}/h]^{-3}$. The solid blue line gives a Gaussian fit to the velocity field
\begin{equation}
P(v)=\frac{1}{\sqrt{2\pi \sigma_v^2}}e^{-\frac{v^2}{2\sigma_v^2}}\, .
\end{equation}
Despite the existence of assembly bias and the fact that the small scale velocity field involves highly non-linear processes, the field can still be reasonably well described by a Gaussian fit except the extreme high-speed wings. The velocity dispersion from the Gaussian fit is $\sigma_v=331{\rm km/s}$.   

Figure~\ref{sys} shows the systematic errors on the measured Hubble constant due to the peculiar velocity field using Eq.~(\ref{H0measure}). From Eq.~(\ref{H0measure}), in addition to the peculiar velocity field, the systematic errors depend also on the Hubble flow redshift $z_{cos}$, which is a combination of the Hubble constant and the gravitational luminosity distant $z_{cos}=H_0^{cos}D_{cos}/c$. Different colors represent the results for GW host galaxies with different Hubble flow redshifts $z_{cos}$. The solid lines represent Gaussian fits to the systematic errors. From Fig.~\ref{sys}, the relative errors drop rapidly with the increasing value of $z_{cos}$, which can be below $1\%$ for GW host galaxies with $z_{cos}>0.13$. 

\section{Conclusions} To summarize, in this work, we have proposed a forward-modeling approach to accurately measure the underlining peculiar velocity field of nearby galaxies, which is crucial to measuring the Hubble constant in the future GW standard-siren experiments. Unlike the conventional backward reconstruction which relies on secondary distance indicators and some simplified assumptions such as the linear perturbation theory, our forward-modeling approach utilizes high-resolution DMO simulations to explore the very non-linear regimes of the peculiar velocity field. An advantage of our approach is that it is based on the first principles of gravity and the only assumption made here is that GR-$\Lambda$CDM is the correct cosmological model. However, this assumption is in line with the GW standard sirens where GR is implicitly assumed to be the correct theory of gravity. Our approach can make the measurement of the Hubble constant from GW standard sirens completely independent of other secondary distance indicators, which therefore can avoid potential systematics associated with those distance indicators. Another advantage of our approach is that it is practically applicable. Indeed, we have already demonstrated our approach using the SDSS main galaxy sample. Once the GW host galaxies are within our galaxy catalog, our results can be directly applied. Our approach can also be naturally extended to future galaxy surveys, such as the dark energy spectroscopic instrument (DESI) survey~\cite{Levi:2013gra}, in particular, the bright time survey (BGS). The BGS of DESI covers an area of $14000\,{\rm deg}^2$, twice as large as the SDSS main galaxy sample. The survey also has a deeper $r$-band magnitude limit $r\sim19.5$, two magnitudes deeper than that of the SDSS main galaxy survey, from which the volume-limited galaxy samples can reach $z>0.1$ even for samples with very high number densities. With such a large volume of galaxies and in combination with the number of detectable GW events in the future GW experiments, it can produce an unprecedented measurement of the Hubble constant ($<1\%$ at a $1\sigma$ level) from nearby galaxies.

\section*{Acknowledgement} 
J.H.H. acknowledges support of Nanjing University and this work used the DiRAC@Durham facility managed by the Institute for Computational Cosmology on behalf of the STFC DiRAC HPC Facility (www.dirac.ac.uk). The equipment was funded by BEIS capital funding via STFC capital grants ST/K00042X/1, ST/P002293/1, ST/R002371/1 and ST/S002502/1, Durham University and STFC operations grant ST/R000832/1. DiRAC is part of the National e-Infrastructure.

\bibliographystyle{unsrt}
\bibliography{myref}

\end{document}